\documentclass[article,shortnames,nojss]{jss}

\usepackage{amsmath, amssymb, algorithm, algorithmic, booktabs, multicol, soul}
\DeclareMathOperator{\pa}{p\hspace{-0.025em}a}
\DeclareMathOperator{\ch}{c\hspace{-0.025em}h}

\newcommand{\dnorm}{\text{N}}

\newcommand{\dbin}{\text{Bin}}

\newcommand{\logit}{\text{logit}}

\newcommand{\distributedas}{\sim}
\newcommand{\given}{\mid}

\newcommand{\stochastic}{S}
\newcommand{\indexederror}{\varepsilon}
\newcommand{\nonindexederror}{\eta}
\newcommand{\deepest}{h^{\star}}

\usepackage{tikz}
\usetikzlibrary{calc,fit,positioning,shapes,arrows}
\tikzset{>={stealth}}

\usepackage{tcolorbox,menukeys}

\usepackage{etoolbox}
\newtoggle{winbugs}
\togglefalse{winbugs}
\newtoggle{openbugs}
\toggletrue{openbugs}

\newcommand{\window}[1]{\textsf{#1}}
\newcommand{\button}[1]{\menu[,]{#1}}
\newcommand{\inputbox}[1]{\textsf{#1}}

\author{Robert J. B. Goudie\\MRC Biostatistics Unit\\University of Cambridge \And
Rebecca M. Turner\\MRC Clinical Trials Unit\\University College London \And
Daniela De Angelis\\MRC Biostatistics Unit\\University of Cambridge \AND
Andrew Thomas\\MRC Biostatistics Unit\\University of Cambridge
}
\title{\proglang{MultiBUGS}: A Parallel Implementation of the \proglang{BUGS} Modelling Framework for Faster Bayesian Inference}

\Plainauthor{Robert J. B. Goudie, Rebecca M. Turner, Daniela De Angelis, Andrew Thomas} %% comma-separated
\Plaintitle{MultiBUGS: A Parallel Implementation of the BUGS Modelling Framework for Faster Bayesian Inference} %% without formatting
\Shorttitle{MultiBUGS: A Parallel Implementation of BUGS for Faster Bayesian Inference} %% a short title (if necessary)

\Abstract{\proglang{MultiBUGS} is a new version of the general-purpose Bayesian modelling software \proglang{BUGS} that implements a generic algorithm for parallelising Markov chain Monte Carlo (MCMC) algorithms to speed up posterior inference of Bayesian models.
The algorithm parallelises evaluation of the product-form likelihoods formed when a parameter has many children in the directed acyclic graph (DAG) representation; and parallelises sampling of conditionally-independent sets of parameters.
A heuristic algorithm is used to decide which approach to use for each parameter and to apportion computation across computational cores.
This enables \proglang{MultiBUGS} to automatically parallelise the broad range of statistical models that can be fitted using \proglang{BUGS}-language software, making the dramatic speed-ups of modern multi-core computing accessible to applied statisticians, without requiring any experience of parallel programming.
We demonstrate the use of \proglang{MultiBUGS} on simulated data designed to mimic a hierarchical e-health linked-data study of methadone prescriptions including 425,112 observations and 20,426 random effects.
Posterior inference for the e-health model takes several hours in existing software, but \proglang{MultiBUGS} can perform inference in only 28 minutes using 48 computational cores.
}
\Keywords{\proglang{BUGS}, parallel computing, Markov chain Monte Carlo, Gibbs sampling, Bayesian analysis, hierarchical models, directed acyclic graph}
\Plainkeywords{BUGS, parallel computing, Markov chain Monte Carlo, Bayesian analysis, hierarchical models, directed acyclic graph}
\Address{
Andrew Thomas\\
MRC Biostatistics Unit\\
School of Clinical Medicine\\
University of Cambridge\\
UK\\
  E-mail: \email{andrew.thomas@mrc-bsu.cam.ac.uk}%\\
}
\begin{document}

\section{Introduction}
\label{sec:intro}
\proglang{BUGS} is a long running project that makes easy to use Bayesian modelling software available to the statistics community.
The software has evolved through three main versions since nineteen eighty-nine: first \proglang{ClassicBUGS} \citep{BUGS:1996}, then \proglang{WinBUGS} \citep{Lunn:2000jc}, then the current open-source \proglang{OpenBUGS} \citep{Lunn:2009jb}.
The software is structured around the twin ideas of the declarative \proglang{BUGS} language \citep{Thomas:2006ta}, through which the user specifies the graphical model \citep{Lauritzen:1990ge} that defines the statistical model to be analysed; and Markov Chain Monte Carlo simulation (MCMC) \citep{Geman:1984hh, Gelfand:1990uc}, which is used to estimate the posterior distribution.
These ideas have also been widely adopted in other Bayesian software, notably in \proglang{JAGS} \citep{JAGSsoftware} and \proglang{NIMBLE} \citep{deValpine:2017hm}, and related ideas are used in  \proglang{Stan} \citep{Carpenter:2017ke}.

Technological advances in recent years have led to massive increases in the amount of data that are generated and stored.
This has posed problems for traditional Bayesian modelling, because fitting such models with a huge amount of data in existing standard software, such as \proglang{OpenBUGS}, is typically either impossible or extremely time-consuming.
While most recent computers have multiple computational cores, which can be used to speed up computation, \proglang{OpenBUGS} has not previously made use of this facility.
The aim of \proglang{MultiBUGS} is to make available to applied statistics practitioners the dramatic speed-ups of multi-core computation without requiring any knowledge of parallel programming, through an easy-to-use implementation of a generic, automatic algorithm for parallelising the MCMC algorithms used by \proglang{BUGS}-style software.

\subsection{Approaches to MCMC parallelisation}

The most straightforward approach for using multiple computational cores or multiple central processing units (CPUs) to perform MCMC simulation is to run each of multiple, independent MCMC chains on a separate CPU or core \citep[e.g.,][]{Bradford:1996bm, Rosenthal:2000tr}.
Since the chains are independent, there is no need for information to be passed between the chains: the algorithm is embarrassingly parallel.
Running several MCMC chains is valuable for detecting problems of non-convergence of the algorithm using, for example, the Brooks-Gelman-Rubin diagnostic \citep{Gelman:1992ts, Brooks:1998um}.
However, the time taken to get past the burn in period cannot be shortened using this approach.

A different approach is to use multiple CPUs or cores for a single MCMC chain, with the aim of shortening the time taken for the MCMC chain to converge and to mix.
One way to do this is to identify tasks within standard MCMC algorithms that can be calculated in parallel, without altering the underlying Markov chain.
A task that is often, in principle, straightforward to parallelise, and is fundamental in several MCMC algorithms, such as the Metropolis-Hastings algorithm, is evaluation of the likelihood \citep[e.g.,][]{Whiley:2004bm, Jewell:2009p10363, Bottolo:2013fg}.
Another task that can be parallelised is sampling of conditionally-independent components, as suggested by, for example, \cite{WilkinsonHandbook:2006}.

\proglang{MultiBUGS} implements all of the above strategies for parallelisation of MCMC.
There are thus two levels of parallelisation: multiple MCMC chains are run in parallel, with the computation required by each chain also parallelised by identifying both complex parallelisable likelihoods and conditionally-independent components that can be sampled in parallel.

There are numerous other approaches to MCMC parallelisation.
Several authors have proposed running parts of the model on separate cores and then combining results \citep{Scott:2013wza} using either somewhat ad hoc procedures or sequential Monte Carlo-inspired methods \citep{Goudie:2016vt}.
This approach has the advantage of being able to reuse already written MCMC software and, in this sense, is similar to the approach used in \proglang{MultiBUGS}.
A separate body of work \citep{Brockwell:2006hd, Angelino:2014tl} proposes using a modified version of the Metropolis-Hastings algorithm which speculatively considers a possible sequence of MCMC steps and evaluates the likelihood at each proposal on a separate core.
The time saving tends to scale logarithmically in the number of cores for this class of algorithms.
A final group of approaches modifies the Metropolis-Hastings algorithm by proposing a sequence of candidate points in parallel \citep{Calderhead:2014hn}.
This approach can reduce autocorrelations in the MCMC chain and so speed up MCMC convergence.

\subsection[MultiBUGS software]{\proglang{MultiBUGS} software}
\proglang{MultiBUGS} is available as free software, under the GNU General Public License version 3, and can be downloaded from \url{https://www.multibugs.org}.
\proglang{MultiBUGS} currently requires Microsoft Windows, and version 8.1 or newer of the Microsoft MPI (MS-MPI) parallel programming framework, available from \url{https://msdn.microsoft.com/en-us/library/bb524831(v=vs.85).aspx}.
Note that the Windows Firewall may require you to give \proglang{MultiBUGS} permission to communicate between cores.
The source code for \proglang{MultiBUGS} can be downloaded from \url{https://github.com/MultiBUGS/MultiBUGS}.
The data and model files to replicate all the results presented in this paper can be found within \proglang{MultiBUGS}, as we describe later in the paper, or can be downloaded from \url{https://github.com/MultiBUGS/multibugs-examples}.

The paper is organised as follows: in Section~\ref{sec:notation} we introduce the class of models we consider and the parallelisation strategy adopted in \proglang{MultiBUGS}; implementation details are provided in Section~\ref{sec:multibugs}; Section~\ref{sec:basic-usage} summarises the basic process of fitting models in \proglang{MultiBUGS}; Section~\ref{sec:example} demonstrates  \proglang{MultiBUGS} for analysing a large hierarchical dataset; and we conclude with a discussion in Section~\ref{sec:conclusions}.

\section{Background and methods}
\label{sec:notation}

\subsection{Models and notation}
\proglang{MultiBUGS} performs inference for Bayesian models that can be represented by a directed acyclic graph (DAG), with each component of the model associated with a node in the DAG.
A DAG \(G = \left( {{V_G},{E_G}} \right)\) consists of a set of nodes or vertices \({V_G}\) joined by directed edges \({E_G} \subset {V_G} \times {V_G}\), represented by arrows.
The parents \(\pa_G(v) = \{u : (u,v) \in E_G\}\) of a node \(v\) are the nodes with an edge pointing to node \(v\).
The children \(\ch_G(v) = \left\{u: (v, u) \in E_G\right\}\) of a node \(v\) are the nodes pointed to by edges emanating from node \(v\).
We omit \(G\) subscripts here, and throughout the paper, wherever there is no ambiguity.

DAGs can be presented graphically (see Figures~\ref{fig:seeds-dag} and \ref{fig:ehealth-dag} below), with stochastic nodes shown in ovals, and constant and observed quantities in rectangles.
Stochastic dependencies are represented by arrows.
Repeated nodes are enclosed by a rounded rectangle (plate), with the range of repetition indicated by the label.

To establish ideas, consider a simple random effects logistic regression model (called ``seeds'') for the number \({r_i}\) of seeds that germinated out of \({n_i}\) planted, in each of \(i = 1, \dots, N = 21\) experiments, with binary indicators of seed type \({X_{1i}}\) and root extract type \({X_{2i}}\) \citep{Crowder:1978ci, Breslow:1993kd}.
\begin{align*}
\begin{split}
r_i &\distributedas \dbin(p_i, n_i)\\
\logit(p_i) &= \alpha_0 + \alpha_1 X_{1i} + \alpha_2 X_{2i} + \alpha_{12} X_{1i} X_{2i} + \beta_i\\
\beta_i &\distributedas \dnorm(\mu_{\beta}, \sigma_{\beta}^{2})
\end{split}
\end{align*}
We choose normal priors for the regression parameters \(\alpha_0, \alpha_1, \alpha_2, \alpha_{12}\), with mean \(\mu_\alpha\) = 0 and standard deviation \(\sigma_\alpha\) = 1000.
We fix \(\mu_{\beta} = 0\), and choose a uniform prior on the range \(\sigma_{\text{min}}  = 0\) to \(\sigma_{\text{max}} = 10\) for the standard deviation \(\sigma_\beta\) of the random effects \(\beta_i\).
Figure~\ref{fig:seeds-dag} shows a DAG representation of the ``seeds'' model.
The data are presented in \cite{Crowder:1978ci}.

\begin{figure}[t]
\centering
\begin{tikzpicture}[minimum width=0.8cm, minimum height = 0.8cm, inner sep = 0.075cm]
\node[draw] (r) at (0, -0.75) {\(r_{i}\)};
\node[draw] (n) at (4.5, 1.5) {\(n_{i}\)};

\node[ellipse, draw] (b) at (0, 1.5) {\(\beta_{i}\)};

\node[draw] (x1) at (1.5, 1.5) {\(X_{1i}\)};
\node[draw] (x2) at (3, 1.5) {\(X_{2i}\)};

\node[ellipse, draw] (sigmabeta) at (1.5, 3.8) {\(\sigma_{\beta}\)};
\node[draw] (mubeta) at (0, 3.8) {\(\mu_{\beta}\)};
\node[ellipse, draw] (alpha0) at (-6, 1.5) {\(\alpha_{0}\)};
\node[ellipse, draw] (alpha1) at (-4.5, 1.5) {\(\alpha_{1}\)};
\node[ellipse, draw] (alpha2) at (-3, 1.5) {\(\alpha_{2}\)};
\node[ellipse, draw] (alpha12) at (-1.5, 1.5) {\(\alpha_{12}\)};

\node[draw] (sigmamin) at (1.5, 5.3) {\(\sigma_{\text{min}}\)};
\node[draw] (sigmamax) at (3, 5.3) {\(\sigma_{\text{max}}\)};
\node[draw] (mu) at (-4.5, 3.8) {\(\mu_{\alpha}\)};
\node[draw] (sigma) at (-3, 3.8) {\(\sigma_{\alpha}\)};

\draw[->] (n) -- (r);

\draw[->] (b) -- (r);
\draw[->] (x1) -- (r);
\draw[->] (x2) -- (r);
\draw[->] (alpha0) -- (r);
\draw[->] (alpha1) -- (r);
\draw[->] (alpha2) -- (r);
\draw[->] (alpha12) -- (r);

\draw[->] (mubeta) -- (b);
\draw[->] (sigmabeta) -- (b);

\draw[->] (sigmamin) -- (sigmabeta);
\draw[->] (sigmamax) -- (sigmabeta);
\draw[->] (mu) -- (alpha0);
\draw[->] (sigma) -- (alpha0);
\draw[->] (mu) -- (alpha1);
\draw[->] (sigma) -- (alpha1);
\draw[->] (mu) -- (alpha2);
\draw[->] (sigma) -- (alpha2);
\draw[->] (mu) -- (alpha12);
\draw[->] (sigma) -- (alpha12);

\node (dummy) at (-0.1, 1.6) {};
\node (dummy2) at (4.6, -0.85) {};

\node[draw, rectangle, rounded corners,
      fit={(dummy) (dummy2) (r) (n) (b) (x1) (x2)}] (plate) {};
\node[font = \footnotesize, node distance=0, inner sep=0pt, below left=-20pt and 5pt of plate.south east] {\(i = 1, \dots, 21\)};
\end{tikzpicture}
\caption{
DAG representation of the seeds model.
}
\label{fig:seeds-dag}
\end{figure}
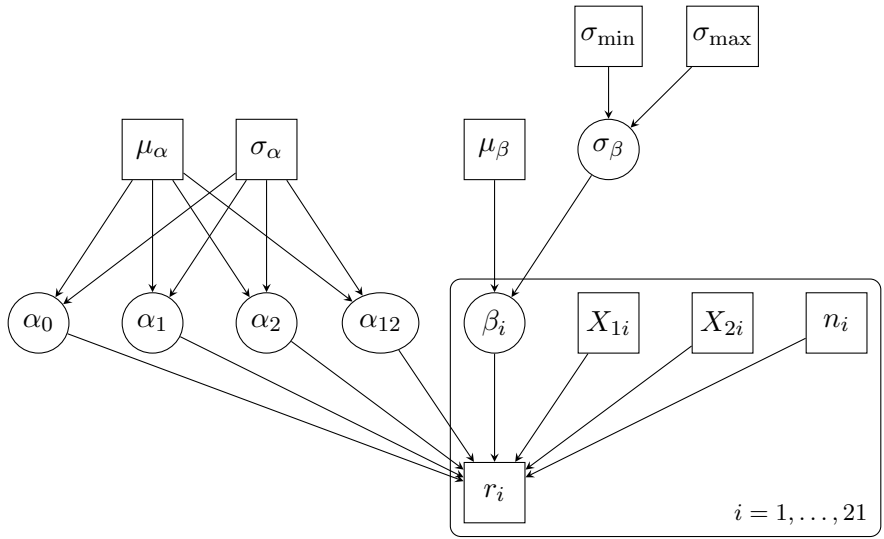

For ease of exposition of the parallelisation methods used by \proglang{MultiBUGS}, we assume throughout this paper that the set of nodes \(V_G\) includes all stochastic parameters \(\stochastic_{G} \subseteq V_{G}\) and constant quantities (including observations and hyperparameters) in the model, but excludes parameters that are entirely determined by other parameters.
As a consequence, the DAG for the seeds example (Figure~\ref{fig:seeds-dag}) includes as nodes the stochastic parameters \(\stochastic_{G} = \{\alpha_0, \allowbreak\alpha_1,\allowbreak \alpha_2, \alpha_{12}, \beta_1, \dots, \beta_{21}, \sigma_{\beta}\}\), the observations \(\{r_i, X_{1i}, X_{2i}, n_i : i = 1, \dots, 21\}\) and the constant hyperparameters \(\{\mu_{\alpha}, \sigma_{\alpha}, \mu_{\beta}, \sigma_{\text{min}}, \sigma_{\text{max}}\}\), but not the parameters that are deterministic functions of other parameters (the germination probabilities \({p_i})\), which have been assimilated into the definition of the distribution of \({r_i}\) before forming the DAG.
Arbitrary DAG models can nevertheless be considered by assimilating deterministic intermediary quantities, such as linear predictors in generalised linear models, into the definition of the conditional distribution of the appropriate descendant stochastic parameter; and considering deterministic prediction separately from the main MCMC computation.
For example, in the seeds example, the random effect precision \(\tau_\beta = \sigma_{\beta}^{-2}\) is deterministically related to the standard deviation \(\sigma_\beta\), so it would not be considered part of the graph if it were of interest: posterior inference for \(\tau_\beta\) could instead be made either by updating its value in the usual (serial) manner after each MCMC iteration, or by post-processing the MCMC samples for \(\sigma_\beta\).

In DAG models, the conditional independence assumptions represented by the DAG mean that the full joint distribution of all quantities \(V\) has a simple factorisation in terms of the conditional distribution \(p(v \given \pa(v))\) of each node \(v \in V\) given its parents \(\pa(v)\):
\begin{equation*}
p(V) = \prod_{v \in V} p(v \given \pa(v))
\end{equation*}
Posterior inference is performed in \proglang{MultiBUGS} by an MCMC algorithm, constructed by associating each node with a suitable updating algorithm, chosen automatically by the program according to the structure of the model.
Most MCMC algorithms involve evaluation of the conditional distribution of the stochastic parameters \(\stochastic \subseteq V\) (at particular values of its arguments).
The conditional distribution \(p(v \given V_{-v})\) of a node \(v \in \stochastic\), given the remaining nodes \({V_{- v}} = V \setminus \{v\}\) is
\begin{equation}
\label{eqn:conditional-distribution}
p(v \given V_{-v})
\propto
p(v \given \pa(v))L(v),
\end{equation}
where \(p(v \given \pa(v))\) is the prior factor and \(L(v) = \prod_{u \in \ch(v)} p(u \given \pa(u))\) is the likelihood factor.

\subsection[Parallelisation methods in MultiBUGS]{Parallelisation methods in \proglang{MultiBUGS}}
\label{sec:methods}

\proglang{MultiBUGS} performs in parallel both multiple chains and the computation required for a single MCMC chain.
In this section, we describe how the computation for a single MCMC chain can be performed in parallel.

\subsubsection{Parallelisation strategies}
MCMC entails sampling, which often requires evaluation of the conditional distribution of the stochastic parameters \(\stochastic\) in the model.
\proglang{MultiBUGS} parallelises these computations for a single MCMC chain via two distinct approaches.

First, when a parameter has many children, evaluation of the conditional distribution is computationally expensive, since Equation~\ref{eqn:conditional-distribution} is the product of many terms. However, the evaluation of the likelihood factor \(L(v)\) can easily be split between \(C\) cores by calculating a partial product involving every \(C\)\textsuperscript{th} child on each core.
With a partition \(\{\ch^{(1)}(v), \dots, \ch^{(C)}(v)\}\) of the set of children \(\ch(v)\), we can evaluate \(\prod_{u \in \ch^{(c)}(v)} p(u \given \pa(u))\) on the \(c\)th core, \(c = 1, \dots, C\).
The prior factor \(p(v \given \pa(v))\) and these partial products can be multiplied together to recover the complete conditional distribution.

Second, when a model includes a large number of parameters then computation may be slow in aggregate, even if sampling of each individual parameter is fast.
However, parameters can clearly be sampled in parallel if they are conditionally independent.
Specifically, all parameters in a set \(W \subseteq \stochastic\) can be sampled in parallel whenever the parameters in \(W\) are mutually conditionally-independent; i.e., all \({w_1} \in
W\) and \({w_2} \in W\) (\(w_1 \neq w_2\)) are conditionally independent given \(V \setminus W\).
If \(C\) cores are available and \(|W|\) denotes the number of elements in the set \(W\), then in a parallel scheme at most \(\lceil |W|/C\rceil\) parameters need be sampled on a core (where \(\lceil x\rceil\) denotes the ceiling function), rather than \(|W|\) in the standard serial scheme.

To identify sets of conditionally-independent parameters, \proglang{MultiBUGS} first partitions the stochastic parameters \(\stochastic\) into depth sets \(D_G^h =\allowbreak \{v \in \stochastic: \allowbreak d_G(v) = h\}\), defined as the set of stochastic nodes with topological depth \(d_G(v) = h\), where topological depth of a node \(v \in V\) is defined recursively, starting from the nodes with no parents.
\begin{equation*}
d_G(v) = \begin{cases}
0 & \text{if } \pa_G(v) = \varnothing \\
1 + \max_{u \in \pa_G(v)} d_G(u) & \text{otherwise}
\end{cases}
\end{equation*}
Note that stochastic nodes \(v \in \stochastic\) have topological depth \(d_G(v) \geq 1\), since the constant hyperparameters of  stochastic nodes are included in the DAG.

Sets of conditionally-independent parameters within a depth set can be identified by noting that all parameters in a set \(W \subseteq D_{G}^{h}\) are mutually conditionally-independent, given the other nodes \(V \setminus W\), if the parameters in \(W\) have no child node in common.
This follows from the \(d\)-separation criterion \citep[Definition 1.2.3,][]{Pearl09}: all such pairs of parameters \({w_1} \in W\) and \({w_2} \in W\) (\(w_1 \neq w_2\)) are \(d\)-separated by \(V \setminus W\) because no `chain path' can exist between \(w_1\) and \(w_2\) because these nodes have the same topological depth; and all `fork paths' are blocked by \(V \setminus W\), as are all `collider paths', except those involving a common child of \(w_1\) and \(w_2\), which are prevented by definition of \(W\).

\subsubsection{Heuristic for determining parallelisation strategy}
A heuristic criterion is used by \proglang{MultiBUGS} to decide which type of parallelism to exploit for each parameter in the model.
The heuristic aims to parallelise the evaluation of conditional distributions of `fixed effect'-like parameters, and parallelise the sampling of `random effect'-like parameters.
The former tend to have a large number of children, whereas the latter tend to have a small number of children.
Each depth set is considered in turn, starting with the `deepest' set \(D_G^{\deepest}\) with \(\deepest = \max_{v \in \stochastic}d_{G}(v)\).
The computation of the parameter's conditional distribution is parallelised if a parameter has more children than double the mean number of children \(\overline{\ch} = \text{mean}_{v \in \stochastic}|\ch_{G}(v)|\), or if all parameters in the graph have topological depth \(h = 1\); otherwise the sampling of conditionally independent sets of parameters is parallelised whenever this is permitted.
The special case for \(h = 1\)  ensures that evaluation of the conditional distribution of parameters is parallelised in `flat' models in which all parameters have identical topological depth.
When a group of parameters is sampled in parallel we would like the time taken to sample each one to be similar, so \proglang{MultiBUGS} assigns parameters to cores in order of the number of children that each parameter has.

\proglang{MultiBUGS} creates a \(C\)-column computation schedule table \(T\), which specifies the parallelisation scheme: where different parameters appear in a row, the corresponding parameters are sampled in parallel; where a single parameter is repeated across a full row, the evaluation of the conditional distribution for that parameter is split into partial products across the \(C\) cores.
A single MCMC iteration consists of evaluating updates as specified by each row of the computation schedule in turn.
The computation schedule includes blanks whenever a set \(W\) of mutually conditionally-independent parameters does not divide equally across the \(C\) cores; that is, when \(|W| \bmod{C} \neq 0\), where \(\bmod{}\) denotes the modulo operator.
The corresponding cores are idle when a blank occurs.
Appendix~\ref{sec:technical-details} describes the algorithms used to create the \(C\)-column computation schedule table \(T\) in detail.

We illustrate the heuristic by describing the process of creating Table~\ref{tab:seeds-schedule}, the computation schedule for the seeds example introduced in Section~\ref{sec:notation}, assuming \(C = 4\) cores are available.
The model includes 26 stochastic parameters \(\stochastic = \{\alpha_0,\allowbreak \alpha_1,\allowbreak \alpha_2,\allowbreak \alpha_{12},\allowbreak \beta_1, \dots, \beta_{21},\allowbreak \allowbreak\sigma_{\beta}\}\); and \(|\ch(\alpha_0)| = |\ch(\alpha_1)| = |\ch(\alpha_2)| = |\ch(\alpha_{12})| = |\ch(\sigma_\beta)| = 21\) and \(|\ch(\beta_1)| = \dots = |\ch(\beta_{21})| = 1\).
\begin{table}
\centering
\begin{tabular}{lcccc}
  \toprule
  & \multicolumn{4}{c}{Core}\\
\cmidrule{2-5}
Row & 1 & 2 & 3 & 4\\
\toprule
1 & \(\beta_1\) & \(\beta_2\) & \(\beta_3\) & \(\beta_4\) \\
\midrule
2 & \(\beta_5\) & \(\beta_6\) & \(\beta_7\) & \(\beta_8\) \\
\midrule
3 &\(\beta_9\) & \(\beta_{10}\) & \(\beta_{11}\) & \(\beta_{12}\) \\
\midrule
4 & \(\beta_{13}\) & \(\beta_{14}\) & \(\beta_{15}\) & \(\beta_{16}\) \\
\midrule
5 & \(\beta_{17}\) & \(\beta_{18}\) & \(\beta_{19}\) & \(\beta_{20}\) \\
\midrule
6 & \(\beta_{21}\) & && \\
\midrule
7 & \(\alpha_{12}\) & \(\alpha_{12}\) & \(\alpha_{12}\) & \(\alpha_{12}\) \\
\midrule
8 & \(\alpha_{1}\) & \(\alpha_{1}\) & \(\alpha_{1}\) & \(\alpha_{1}\) \\
\midrule
9 & \(\alpha_{2}\) & \(\alpha_{2}\) & \(\alpha_{2}\) & \(\alpha_{2}\) \\
\midrule
10 & \(\alpha_{0}\) & \(\alpha_{0}\) & \(\alpha_{0}\) & \(\alpha_{0}\) \\
\midrule
11 & \(\sigma_{\beta}\) & \(\sigma_{\beta}\) & \(\sigma_{\beta}\) & \(\sigma_{\beta}\) \\
\bottomrule
\end{tabular}
\caption{Computation schedule table \(T\) for the seeds example, with 4 cores.}
\label{tab:seeds-schedule}
\end{table}
\proglang{MultiBUGS} first considers the parameters \(\beta_1, \dots, \beta_{21}\), since the topological depth \(d(\beta_1) = \dots = d(\beta_{21}) = 2 = \max_{v \in \stochastic}d(v)\).
None of the likelihood evaluation for \(\beta_1, \dots, \beta_{21}\) is parallelised, because all these parameters have only 1 child and \(\overline{\ch} \approx 4.8\).
However, \(\beta_1, \dots, \beta_{21}\) are mutually conditionally-independent and so these parameters are distributed across the 4 cores as shown in the first 6 rows of Table~\ref{tab:seeds-schedule}. Since \(21 \bmod{4} \neq 0\), cores 2, 3 and 4 will be idle while \(\beta_{21}\) is sampled.
Next, we consider \(\alpha_0, \alpha_1, \alpha_2, \alpha_{12}\) and \(\sigma_\beta\), since \(d(\alpha_0) = \dots = d(\alpha_{12}) = d(\sigma_\beta) = 1\).
Since all of these parameters have 21 children and \(\overline{\ch}  \approx 4.8\), \proglang{MultiBUGS} will spread the likelihood evaluation of all these parameters across cores, and these are assigned to the computation schedule in turn.

\subsubsection{Block samplers}
\proglang{MultiBUGS} is able to use a block MCMC sampler when appropriate: that is, algorithms that sample a block of nodes jointly, rather than just a single node at a time.
Block samplers are particularly beneficial when parameters in the model are highly correlated \textit{a posteriori} \cite[see e.g.,][]{Roberts:1997cb}.
The conditional distribution for a block \(B \subseteq \stochastic\) of nodes, given the rest of nodes \(V_{-B} = V \setminus B\), is
\begin{equation*}
p(B \given V_{-B})
\propto
\prod_{b \in B} p(b \given \pa(b))
\times
\prod_{b \in B}
\prod_{u \in \ch(b)}
p(u \given \pa(u))
\end{equation*}
Block samplers can be parallelised in a straightforward manner: if we consider a block \(B\) as a single node, and define \(\ch(B) = \cup_{b \in B} \ch(b)\), then the approach introduced above is immediately applicable, and we can exploit both opportunities for parallelisation for block updates.
A mixture of single node and block updaters can be used without complication.

In the seeds example it is possible to block together \(\alpha_0, \alpha_1, \alpha_2, \alpha_{12}\).
The block then has 21 children, and so our algorithm chooses to spread evaluation of their likelihood over multiple cores.
The computation schedule remains identical to Table~\ref{tab:seeds-schedule}, but the block sampler waits until all the likelihoods corresponding to rows 7 to 10 of Table~\ref{tab:seeds-schedule} are evaluated before determining each update for the \(\{\alpha_0, \alpha_1, \alpha_2, \alpha_{12}\}\) block.

\section{Implementation details}
\label{sec:multibugs}

\proglang{BUGS} represents statistical models internally using a dynamic object-oriented data structure \citep{Warford02} that is analogous to a DAG.
The nodes of the graph are objects and the edges of the graph are pointers contained in these objects.
Although the graph is specified in terms of the parents of each node, \proglang{BUGS} identifies the children of each node and stores this as a list embedded in each parameter node.
Each node object has a value and a method to calculate its probability density function.
For observations and fixed hyperparameters the value is fixed and is read in from a data file; for parameters the value is variable and is sampled within a MCMC algorithm.
Each MCMC sampling algorithm is represented by a class \citep{Warford02} and a new sampling object of an appropriate class is created for each parameter in the statistical model.
Each sampling object contains a link to the node (or block of nodes) in the graphical model that represents the parameter (or block of parameters) being sampled.
One complete MCMC update of the model involves a traversal of a list of all these sampling objects, with each object's sampling method called in turn.
\cite{Lunn:2000jc} provides further background on the internal design of \proglang{BUGS}.

The \proglang{MultiBUGS} software consists of two distinct computer programs: a user interface and a computational engine.
The computational engine is a small program assembled by linking together some modules of the \proglang{OpenBUGS} software plus a few additional modules to implement our parallelisation algorithm.
Copies of the computational engine run on multiple cores and communicate with each other using the message passing interface (\proglang{MPI}) protocol \citep{Pacheco:1997}, version 2.0.
The user interface program is a slight modification (and extension) of the \proglang{OpenBUGS} software.
The user interface program compiles an executable ``worker program'' that contains the computational engine required for a particular statistical model.
It also writes out a file containing a representation of the data structures that specify the statistical model.
It then starts a number of copies of the computational engine on separate computer cores.
These worker programs then read in the model representation file to rebuild the graphical model and start generating MCMC samples using our distributed algorithms.
The worker programs communicate with the user interface program via an \proglang{MPI} intercommunicator object.
The user interface is responsible for calculating summary statistics of interest.

Both sources of parallelism described in Section~\ref{sec:methods} require only simple modifications of the data structures and algorithms used in the \proglang{BUGS} software.
Each core keeps a copy of the current state of the MCMC, as well as two pseudo-random number generation (PRNG) streams \citep{WilkinsonHandbook:2006}: a ``core-specific'' stream, initialised with a different seed for each core; and ``common'' stream, initialised using the same seed on all cores.
Initially, each core loads the sampling algorithm, the computation schedule, and the complete DAG, which is then altered as follows so that the overall computation yields the computation required for the original, complete DAG.

When the calculation of a parameter's likelihood is parallelised across cores, the list of children associated with a parameter on each core is thinned (pruned) so that it contains only the children in the corresponding partition component of \(\ch(v)\).
The \proglang{BUGS} MCMC sampling algorithm implementations then require only minor changes so that the partial likelihoods are communicated between cores.
For example, a random walk Metropolis algorithm \citep{Metropolis:1953in} is performed as follows: first, on each core, the prior factor and a partial likelihood contributions to the conditional distribution are calculated for the current value of the parameter.
Each core then samples a candidate value.
These candidates will be identical across cores, since the ``common'' PRNG stream is used.
The prior and partial likelihood contributions are then calculated for the candidate value, and the difference between the two partial log-likelihood contributions can be combined across cores using the \proglang{MPI} function \code{Allreduce}.
The usual Metropolis test can then be applied on each core in parallel using the ``common'' PRNG stream, after which the state of Markov chain is identical across cores.
Computation of the prior factor and the Metropolis test is intentionally duplicated on every core because we found that the time taken to evaluate these quantities is usually shorter than the time taken to propagate their result across cores.

When a set of parameters \(W\) is sampled in parallel over the worker cores, the list of MCMC sampling objects is thinned on each core so that only parameters specified by the corresponding column of the computation schedule are updated on each core.
The existing MCMC sampling algorithm implementations used in \proglang{OpenBUGS} can then be used without modification with each ``core-specific'' PRNG stream.
The \proglang{MPI} function \code{Allgather} is used to send newly sampled parameters to each core.
Note we need run \code{Allgather} only after each core has sampled all of its assigned components in \(W\), rather than after each component in \(W\) is sampled.
For example, in the seeds example, we use \code{Allgather} after row 6.
This considerably reduces message-passing overheads when the number of elements in \(W\) is large.

Running multiple chains is handled via standard \proglang{MPI} methods.
If we have, say, two chains and eight cores, we partition the cores into two sets of four cores and set up separate \proglang{MPI} collective communicators \citep{Pacheco:1997} for each set of cores for \code{Allreduce} and \code{Allgather} to use.
Requests can be sent from the master to the workers using the intercomunicater and results returned.
We find it useful to designate a special ``lead worker'' for each chain that we simulate.
Each of these lead workers sends back sampled values to the master, where summary statistics can be collected.
Only sampled values corresponding to quantities that the user is monitoring need to be returned to the master.
This can considerably reduce the amount of communication between the workers and the master.

\section[Basic usage of MultiBUGS]{Basic usage of \proglang{MultiBUGS}}
\label{sec:basic-usage}

The procedure for running a model in \proglang{MultiBUGS} is largely the same as in \proglang{WinBUGS} or \proglang{OpenBUGS}.
\proglang{MultiBUGS} adopts the standard \proglang{BUGS} language for specifying models, the core of which is common also to \proglang{WinBUGS}, \proglang{OpenBUGS}, \proglang{JAGS} and \proglang{NIMBLE}.
A detailed tutorial on the use of \proglang{BUGS} can be found in, for example, \cite{Lunn:2013}.

\begin{figure}
\centering
\includegraphics[width=0.3\linewidth]{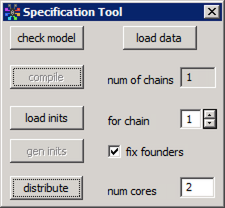}
\caption{The Specification Tool in \proglang{MultiBUGS}, including the `distribute' button, which is used to initialise the parallelisation.}
\label{fig:multibugs-dialog}
\end{figure}
An analysis is specified in \proglang{MultiBUGS} using the \window{Specification Tool} (\menu[,]{Model, Specification...}\,) by checking the syntax of a model (\button{check model}), loading the data (\button{load data}), compiling (\button{compile}) and setting up initial values (\button{load inits} and \button{gen inits}).

We then specify the total number of cores to distribute computation across by entering a number in the box labelled \inputbox{num cores} (Figure~\ref{fig:multibugs-dialog}) and then clicking \button{distribute}.
This should be set at a value less than or equal to the number of processing cores available on your computer (the default is 2).
If multiple chains are run, the cores will be divided equally across chains.
We recommend that users experiment with different numbers of cores, since the setting that leads to fastest computation depends both on the specific model and data being analysed and on the computing hardware being used.
While increased parallelisation will often result in faster computation, in some cases communication overheads will balloon to the point where parallelisation gains are overturned.
Furthermore, Amdahl's bound \citep{Amdahl:1967gg} on the speed-up that is theoretically obtainable with increased parallelisation may also be hit in some settings.
Note that changing the number of cores will alter the exact samples obtained, since this affects the PRNG stream used to draw each sample (as described in Section~\ref{sec:multibugs}).

Samples are drawn using the \window{Update Tool} (\menu[,]{Model, Update...}).
The use of the \window{Sample Monitor Tool} (\menu[,]{Inference, Samples...}\,) to monitor parameters; to assess MCMC convergence, using, for example, the Brooks-Gelman-Rubin diagnostic \citep{Gelman:1992ts, Brooks:1998um}; and to obtain results is the same as in \proglang{WinBUGS} and \proglang{OpenBUGS}.
Analyses can be automated in \proglang{MultiBUGS} using the same simple procedural scripting language that is available in \proglang{OpenBUGS}.
The new command \code{modelDistribute(C)} can be used to specify that parallelisation should be across \code{C} cores; for details see \menu[,]{Manuals, MultiBUGS User Manual, Scripts and Batch-mode}\,.

\subsection{Seeds example}
The model, data and initial conditions for the seeds examples can be found within \proglang{MultiBUGS} in \menu[,]{Manuals, Examples Vol I, Seeds: random effects logistic regression}\,.
This is a simple model involving a small number of parameters and observations, so computation is already fast in \proglang{OpenBUGS} and is no faster in \proglang{MultiBUGS} (both take less than a second to do 1000 MCMC updates) because the benefit of parallelisation is cancelled out by communication overheads.
However, for some more complicated models, \proglang{MultiBUGS} will be dramatically faster than \proglang{OpenBUGS}.
We illustrate this with an example based on e-health data.

\section{Illustration of usage with hierarchical e-health data}
\label{sec:example}
Our e-health example is based on a large linked database of methadone prescriptions given to opioid dependent patients in Scotland, which was used to examine the influence of patient characteristics on doses prescribed \citep{Gao:2016hq, DimitropolouUnpub}.
This example is typical of many databases of linked health information drawn from primary care records, hospital records, prescription data and disease/death registries.
Each data source often has a hierarchical structure, arising from regions, institutions and repeated measurements within individuals.
Here, since we are unable to share the original dataset, we analyse a synthetic dataset, simulated to match the key traits of the original dataset.

The model includes 20,426 random effects in total, and was fitted to 425,112 observations.
It is possible to fit this model using standard MCMC simulation in \proglang{OpenBUGS} but, unsurprisingly, the model runs extremely slowly and it takes 32 hours to perform a sufficient number of iterations (15,000) to satisfy standard convergence assessment diagnostics.
In such data sets it can be tempting to choose a much simpler and faster method of analysis, but this may not allow appropriately for the hierarchical structure or enable exploration of sources of variation.
Instead it is preferable to fit the desired hierarchical model using MCMC simulation, while speeding up computation as much as possible by exploiting parallel processing.

The model code, data and initial conditions can be found within \proglang{MultiBUGS} in \menu[,]{Manuals, Examples Vol IV, Methadone: an E-health model}\,.

\subsection{E-health data}
The data have a hierarchical structure, with multiple prescriptions nested within patients within regions.
For some of the outcome measurements, person identifiers and person-level covariates are available (240,776 observations).
These outcome measurements \({y_{ijk}}\) represent the quantity of methadone prescribed on occasion \(k\) for person \(j\) in region \(i\) (\(i = 1, \dots, 8;\; j = 1, \dots, J_i;\; k = 1, \dots, K_{ij}\)), and are recorded in the file \code{ehealth_data_id_available}.
Each of these measurements is associated with a binary covariate \({r_{ijk}}\) (called \code{source.indexed}) that indicates the source of prescription on occasion \(k\) for person \(j\) in region \(i\), with \(r_{ijk} = 1\) indicating that the prescription was from a General Practitioner (family physician).
No person identifiers or person-level covariates are available for the remaining outcome measurements (184,336 observations).
We denote by \({z_{il}}\) the outcome measurement for the \(l\)\textsuperscript{th} prescription without person identifiers in region \(i\) (\(i = 1, \dots, 8; \; l = 1, \dots, L_i\)).
These data are in the file \code{ehealth_data_id_missing}.
A binary covariate \(s_{il}\) (called \code{source.nonindexed)} indicates the source of the \(l\)\textsuperscript{th} prescription without person identifiers in region \(i\), with \(s_{il} = 1\) indicating that the prescription was from a General Practitioner (family physician).
The final data file, \code{ehealth_data_n}, contains several totals used in the \proglang{BUGS} code.

\subsection{E-health model}
We model the effect of the covariates with a regression model, with regression parameter \(\beta_m\) corresponding to the \(m\)\textsuperscript{th} covariate \(x_{mij}\) (\(m = 1, \dots, 4\)), while allowing for within-region correlation via region-level random effects \({u_i}\), and within-person correlation via person-level random effects \({w_{ij}}\); source effects \({v_i}\) are assumed random across regions.
\begin{align*}
y_{ijk} &= \sum_{m = 1}^{4} \beta_m x_{mij} + u_i + v_i r_{ijk} + w_{ij} + \indexederror_{ijk}\\
u_i \distributedas \dnorm(\mu_u, \sigma^{2}_u), &\; v_i \distributedas \dnorm(\mu_v, \sigma^{2}_v), \; w_{ij} \distributedas \dnorm(\mu_{w}, \sigma^{2}_w), \; \indexederror_{ijk} \distributedas \dnorm(\mu_{\indexederror}, \sigma^{2}_\indexederror)
\end{align*}
The means \(\mu_{w}\) and \(\mu_{\indexederror}\) are both fixed to 0.

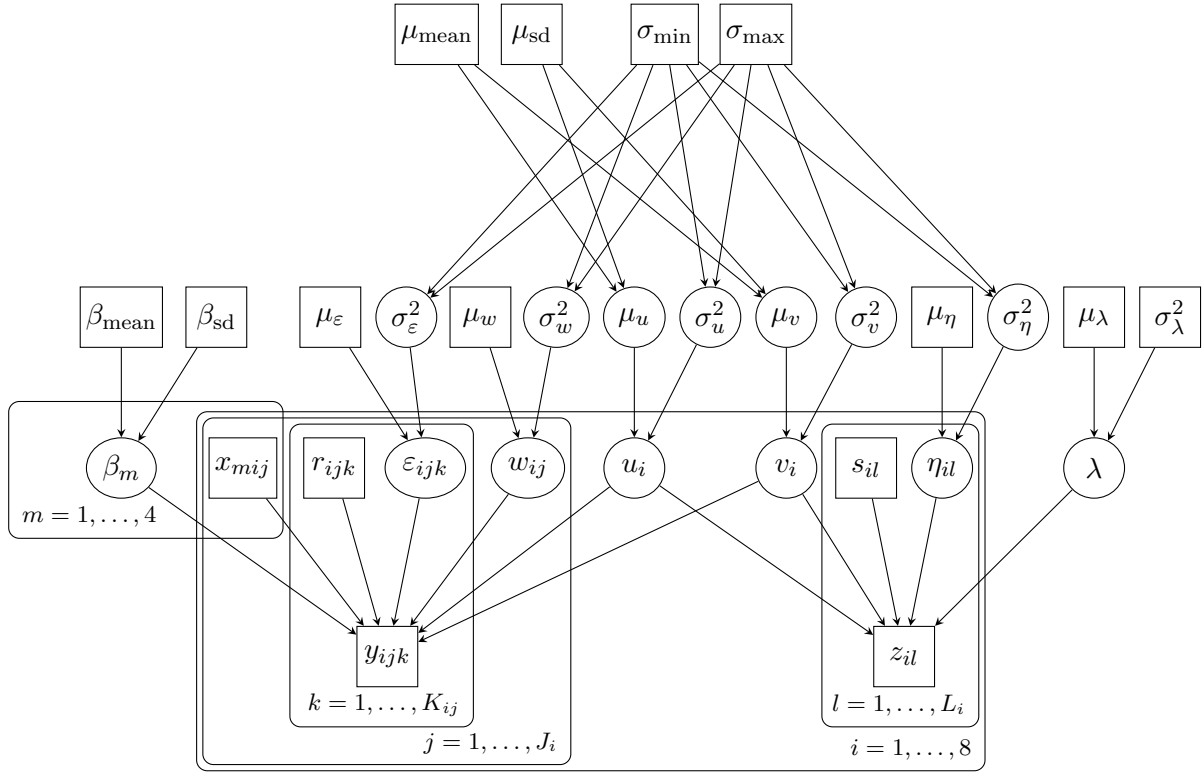
\begin{figure}
\centering
\begin{tikzpicture}[minimum width=0.8cm, minimum height = 0.8cm, inner sep = 0.075cm]
\node[draw] (yijk) at (0.6, -1) {\(y_{ijk}\)};
\node[draw] (zil) at (7.4, -1) {\(z_{il}\)};

\node[ellipse, draw] (betam) at (-2.9, 1.5) {\(\beta_{m}\)};
\node[draw] (xmij) at (-1.3, 1.5) {\(x_{mij}\)};
\node[ellipse, draw] (indexederror) at (1.1, 1.5) {\(\indexederror_{ijk}\)};
\node[draw] (sijk) at (-0.1, 1.5) {\(r_{ijk}\)};
\node[ellipse, draw] (wij) at (2.45, 1.5) {\(w_{ij}\)};
\node[ellipse, draw] (ui) at (3.85, 1.5) {\(u_{i}\)};
\node[ellipse, draw] (vi) at (5.85, 1.5) {\(v_{i}\)};
\node[draw] (sil) at (6.9, 1.5) {\(s_{il}\)};
\node[ellipse, draw] (nonindexederror) at (7.9, 1.5) {\(\nonindexederror_{il}\)};
\node[ellipse, draw] (lambda) at (9.9, 1.5) {\(\lambda\)};

\node[draw] (muindexed) at (-0.15, 3.5) {\(\mu_{\indexederror}\)};
\node[draw] (muw) at (1.825, 3.5) {\(\mu_{w}\)};
\node[draw] (munonindexed) at (7.9, 3.5) {\(\mu_{\nonindexederror}\)};

\node[ellipse, draw] (sigmaindexed) at (0.85, 3.5) {\(\sigma_{\indexederror}^{2}\)};
\node[ellipse, draw] (sigmaw) at (2.825, 3.5) {\(\sigma_{w}^{2}\)};
\node[ellipse, draw] (sigmanonindexed) at (8.9, 3.5) {\(\sigma_{\nonindexederror}^{2}\)};

\node[ellipse, draw] (mu) at (3.85, 3.5) {\(\mu_{u}\)};
\node[ellipse, draw, inner ysep = 0] (sigmau) at (4.85, 3.5) {\(\sigma_{u}^{2}\)};

\node[ellipse, draw] (muv) at (5.85, 3.5) {\(\mu_{v}\)};
\node[ellipse, draw] (sigmav) at (6.9, 3.5) {\(\sigma_{v}^{2}\)};

\node[draw] (sigmamin) at (4.25, 7.25) {\(\sigma_{\text{min}}\)};
\node[draw] (sigmamax) at (5.45, 7.25) {\(\sigma_{\text{max}}\)}; % 3.45

\node[draw] (mumean) at (1.25, 7.25) {\(\mu_{\text{mean}}\)};
\node[draw] (musd) at (2.5, 7.25) {\(\mu_{\text{sd}}\)};

\node[draw] (lambdamean) at (9.9, 3.5) {\(\mu_{\lambda}\)};
\node[draw] (lambdasd) at (10.9, 3.5) {\(\sigma_{\lambda}^{2}\)};

\node[draw] (betamean) at (-2.9, 3.5) {\(\beta_{\text{mean}}\)};
\node[draw] (betasd) at (-1.65, 3.5) {\(\beta_{\text{sd}}\)};

\draw[->] (betam) -- (yijk);
\draw[->] (xmij) -- (yijk);
\draw[->] (indexederror) -- (yijk);
\draw[->] (sijk) -- (yijk);
\draw[->] (wij) -- (yijk);
\draw[->] (ui) -- (yijk);
\draw[->] (vi) -- (yijk);

\draw[->] (sil) -- (zil);
\draw[->] (ui) -- (zil);
\draw[->] (vi) -- (zil);
\draw[->] (nonindexederror) -- (zil);
\draw[->] (lambda) -- (zil);

\draw[->] (muindexed) -- (indexederror);
\draw[->] (muw) -- (wij);
\draw[->] (munonindexed) -- (nonindexederror);
\draw[->] (sigmaindexed) -- (indexederror);
\draw[->] (sigmaw) -- (wij);
\draw[->] (sigmanonindexed) -- (nonindexederror);

\draw[->] (mu) -- (ui);
\draw[->] (sigmau) -- (ui);

\draw[->] (muv) -- (vi);
\draw[->] (sigmav) -- (vi);

\draw[->] (sigmamin) -- (sigmaindexed);
\draw[->] (sigmamin) -- (sigmaw);
\draw[->] (sigmamin) -- (sigmanonindexed);
\draw[->] (sigmamin) -- (sigmau);
\draw[->] (sigmamin) -- (sigmav);

\draw[->] (sigmamax) -- (sigmaindexed);
\draw[->] (sigmamax) -- (sigmaw);
\draw[->] (sigmamax) -- (sigmanonindexed);
\draw[->] (sigmamax) -- (sigmau);
\draw[->] (sigmamax) -- (sigmav);

\draw[->] (mumean) -- (mu);
\draw[->] (mumean) -- (muv);

\draw[->] (musd) -- (mu);
\draw[->] (musd) -- (muv);

\draw[->] (lambdamean) -- (lambda);
\draw[->] (lambdasd) -- (lambda);

\draw[->] (betamean) -- (betam);
\draw[->] (betasd) -- (betam);

\node (dummyk) at (-0.2, -1.45) {};
\node (dummyk2) at (1.25, 1.6) {};
\node (dummyj) at (0, -1.95) {};
\node (dummym) at (-3.9, 1.05) {};
\node (dummym2) at (-3.9, 1.9) {};
\node (dummyl) at (6.8, 1.6) {};
\node (dummyl2) at (6.8, -1.45) {};

\node[draw, rectangle, rounded corners,
      fit={(xmij) (betam) (dummym) (dummym2)} ] (platem) {};
\node[font = \footnotesize, node distance=0, inner sep=0pt, below right=-20pt and 5pt of platem.south west] {\(m = 1, \dots, 4\)};

\node[draw, rectangle, rounded corners,
      fit={(yijk) (indexederror) (sijk) (dummyk) (dummyk2)} ] (platek) {};
\node[font = \footnotesize, node distance=0, inner sep=0pt, below left=-20pt and 5pt of platek.south east] {\(k = 1, \dots, K_{ij}\)};

\node[draw, rectangle, rounded corners,
      fit={(yijk) (indexederror) (xmij) (sijk) (wij) (platek) (dummyj)} ] (platej) {};
\node[font = \footnotesize, node distance=0, inner sep=0pt, below left=-20pt and 5pt of platej.south east] {\(j = 1, \dots, J_{i}\)};

\node[draw, rectangle, rounded corners,
      fit={(zil) (nonindexederror) (dummyl) (dummyl2)} ] (platel) {};
\node[font = \footnotesize, node distance=0, inner sep=0pt, below left=-20pt and 5pt of platel.south east] {\(l = 1, \dots, L_{i}\)};

\node[draw, rectangle, rounded corners,
      fit={(yijk) (zil) (indexederror) (xmij) (sijk) (wij) (ui) (vi) (nonindexederror) (platej) (platel)} ] (platei) {};
\node[font = \footnotesize, node distance=0, inner sep=0pt, below left=-20pt and 5pt of platei.south east] {\(i = 1, \dots, 8\)};

\end{tikzpicture}
\caption{DAG representation of the e-health model.}
\label{fig:ehealth-dag}
\end{figure}

The outcome measurements \({z_{il}}\) contribute only to estimation of regional effects \({u_i}\) and source effects \({v_i}\).
\begin{align*}
z_{il} &= \lambda + u_i + v_i s_{il} + \nonindexederror_{il}\\
\nonindexederror_{il} &\distributedas \dnorm(\mu_{\eta}, \sigma^{2}_\nonindexederror)
\end{align*}
The error variance \(\sigma_\nonindexederror^2\) represents a mixture of between-person and between-occasion variation.
We fix the error mean \(\mu_{\eta} = 0\).
We assume uniform priors for \(\sigma_{u}, \sigma_{v}, \sigma_{w}, \sigma_{\indexederror}, \sigma_{\nonindexederror}\) on the range \(\sigma_{\text{min}} = 0\) to \(\sigma_{\text{max}} = 10\), and normal priors for \(\beta_1, \dots, \beta_4, \mu_u, \mu_v\) and \(\lambda\) with mean \(\beta_{\text{mean}} = \mu_{\text{mean}} = \mu_{\lambda} = 0\) and standard deviation \(\beta_{\text{sd}} = \mu_{\text{sd}} = \mu_{\lambda} = 100\).
Figure~\ref{fig:ehealth-dag} is a DAG representation of this model.

The data have been suitably transformed so that fitting a linear model is appropriate.
We do not consider alternative approaches to analysing the data set.
The key parameters of interest are the regression parameters \(\beta_1, \dots, \beta_4\) and the standard deviations \(\sigma_u\) and \(\sigma_v\) for the region and source random effects.

This model can be specified in \proglang{BUGS} as follows:

\begin{CodeInput}
model {
  # Outcomes with person-level data available
  for (i in 1:n.indexed) {
    outcome.y[i] ~ dnorm(mu.indexed[i], tau.epsilon)
    mu.indexed[i] <- beta[1] * x1[i] +
                     beta[2] * x2[i] +
                     beta[3] * x3[i] +
                     beta[4] * x4[i] +
                     region.effect[region.indexed[i]] +
                     source.effect[region.indexed[i]] * source.indexed[i] +
                     person.effect[person.indexed[i]]
  }

  # Outcomes without person-level data available
  for (i in 1:n.nonindexed){
    outcome.z[i] ~ dnorm(mu.nonindexed[i], tau.eta)
    mu.nonindexed[i] <- lambda +
                        region.effect[region.nonindexed[i]] +
                        source.effect[region.nonindexed[i]] *
                                      source.nonindexed[i]
  }

  # Hierarchical priors
  for (i in 1:n.persons){
    person.effect[i] ~ dnorm(0, tau.person)
  }
  for (i in 1:n.regions) {
    region.effect[i] ~ dnorm(mu.region, tau.region)
    source.effect[i] ~ dnorm(mu.source, tau.source)
  }

  lambda ~ dnorm(0, 0.0001)
  mu.region ~ dnorm(0, 0.0001)
  mu.source ~ dnorm(0, 0.0001)

  # Priors for regression parameters
  for (m in 1:4){
    beta[m] ~ dnorm(0, 0.0001)
  }

  # Priors for variance parameters
  tau.eta <- 1/pow(sd.eta, 2)
  sd.eta ~ dunif(0, 10)
  tau.epsilon <- 1/pow(sd.epsilon, 2)
  sd.epsilon ~ dunif(0, 10)
  tau.person <- 1/pow(sd.person, 2)
  sd.person ~ dunif(0, 10)
  tau.source <- 1/pow(sd.source, 2)
  sd.source ~ dunif(0, 10)
  tau.region <- 1/pow(sd.region, 2)
  sd.region ~ dunif(0, 10)
}
\end{CodeInput}

\subsection{E-health initial values}
For chain 1, we used the following initial values:
\begin{CodeInput}
list(lambda = 0, beta = c(0, 0, 0, 0), mu.source = 0, sd.epsilon = 0.5,
     sd.person = 0.5, sd.source = 0.5, sd.region = 0.5, sd.eta = 0.5)
\end{CodeInput}
and for chain 2 we used:
\begin{CodeInput}
list(lambda = 0.5, beta = c(0.5, 0.5, 0.5, 0.5), mu.source = 0.5,
     sd.epsilon = 1, sd.person = 1, sd.source = 1, sd.region = 1,
     sd.eta = 1)
\end{CodeInput}

\subsection[Parallelisation in MultiBUGS]{Parallelisation in \proglang{MultiBUGS}}
After setting the number of cores, the computation schedule chosen by \proglang{MultiBUGS} can be viewed in \menu[,]{Info,Show distribution}\,.
\proglang{MultiBUGS} parallelises sampling of all the person-level random effects \({w_{ij}}\), except for the component corresponding to the person with the most observations (176 observations); \proglang{MultiBUGS} parallelises likelihood computation of this component instead.
The likelihood computation of all the other parameters in the model is also parallelised, except for the mutually conditionally-independent sets \(\{\mu_u, \mu_v\}\) and \(\{\sigma^{2}_u, \sigma^{2}_v\}\), which are sampled in parallel in turn.

\subsection[Run time comparisons across BUGS implementations]{Run time comparisons across \proglang{BUGS} implementations}
To demonstrate the speed-up possible in \proglang{MultiBUGS} using a range of number of cores, we ran two chains for 15,000 updates for the e-health example.
This run length was chosen to mimic realistic statistical practice, since, after discarding the first 5,000 iterations as burn-in, visual inspection of chain-history plots and the Brooks-Gelman-Rubin diagnostic \citep{Gelman:1992ts, Brooks:1998um} indicated convergence.
We ran the simulations (each replicated three times) on a sixty four core machine consisting of four sixteen-core 2.4Ghz AMD Operon 6378 processors with 128GB shared RAM.

\begin{figure}
\centering
\includegraphics[width=0.45\linewidth]{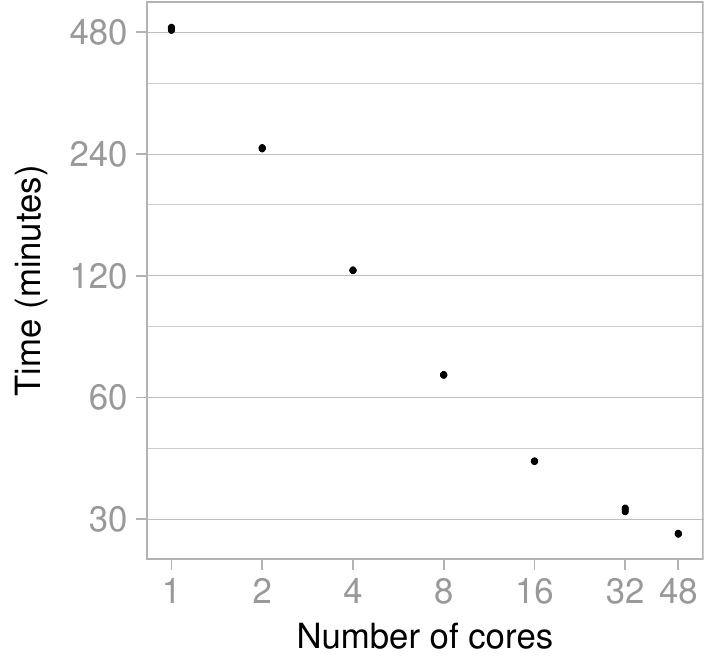}
\caption{
Run time against number of cores for 15,000 iterations of the e-health example model, running 2 chains simultaneously.
The run time in each of 3 replicate runs are shown.
Both time and number of cores are displayed on a \(\log\) scale.
}
\label{fig:runtimes}
\end{figure}

Figure~\ref{fig:runtimes} shows the run time against the number of cores on a log-log scale.
Substantial time savings are achieved using \proglang{MultiBUGS}: on average using one core took 8 hours 10 minutes; using two cores took 4 hours and 8 minutes; and using forty-eight cores took only 28 minutes.
In contrast, these simulations took 32 hours in standard single-core \proglang{OpenBUGS} 3.2.3; and 9 hours using \proglang{JAGS} 4.0.0 via \proglang{R} 3.3.1.

The scaling of performance with increasing number of cores is good up to sixteen cores and then displays diminishing gains.
This may be due to inter core communication being much faster within each processor of 16 cores compared to across processors, or the diminishing returns anticipated by Amdahl's law \citep{Amdahl:1967gg}.
Running only one chain approximately halved the run time for two chains.

\subsection{Results}
The posterior summary table we obtained is as follows:

\begin{minipage}{\textwidth}
{\footnotesize
\begin{CodeOutput}
           mean      median    sd       MC_error  val2.5pc  val97.5pc  start  sample  ESS
beta[1]    -0.07124  -0.07137  0.01272  5.784E-4  -0.09561  -0.0461    5001   20000   483
beta[2]    -0.2562   -0.2563   0.02437  9.186E-4  -0.3036   -0.208     5001   20000   704
beta[3]    0.1308    0.1311    0.0114   5.7E-4    0.1085    0.1528     5001   20000   399
beta[4]    0.13      0.1305    0.0182   7.083E-4  0.09474   0.1651     5001   20000   660
sd.region  1.259     1.157     0.4606   0.005305  0.7024    2.445      5001   20000   7536
sd.source  0.3714    0.3417    0.1359   0.001611  0.2057    0.7153     5001   20000   7116
\end{CodeOutput}
}
\end{minipage}

\section{Discussion}
\label{sec:conclusions}
\proglang{MultiBUGS} makes Bayesian inference using multi-core processing accessible for the first time to applied statisticians working with the broad class of statistical models available in \proglang{BUGS} language software.
It adopts a pragmatic algorithm for parallelising MCMC sampling, which we have demonstrated speeds up inference in a random-effects logistic regression model involving a large number of random effects and observations.
While a large literature has developed proposing methods for parallelising MCMC algorithms (see Section~\ref{sec:intro}), a generic, easy-to-use implementation of these ideas has been heretofore lacking.
Almost all users of \proglang{BUGS} language software will have a multi-core computer available, since desktop computers typically now have a moderate number (up to ten) of cores, and laptops typically have 2-4 cores.
However, workstations with an even larger number of cores are now becoming available: for example, Intel's Xeon Phi x200 processor contains between sixty-four and seventy-two cores.

The magnitude of speed-up provided by \proglang{MultiBUGS} depends on the model and data being analysed and the computer hardware being used.
Two levels of parallelisation can be used in \proglang{MultiBUGS}: independent MCMC chains can be parallelised, and then computation within a single MCMC chain can be parallelised.
The first level of parallelisation will almost always be advantageous whenever sufficient cores are available, since no communication across cores is needed.
The gain from second level of parallelisation is problem specific: the gain will be largest for models involving parameters with a large number of likelihood terms and/or a large number of conditionally independent parameters.
For example, \proglang{MultiBUGS} is able to parallelise inference for many standard regression-type models involving both fixed and random effects, especially with a large number of observations, since fixed effect regression parameters will have a large number of children (the observations), and random effects will typically be conditionally independent.
For models without these features, the overheads of the second level of parallelisation may outweigh the gains on some computing hardware, meaning only the first level of parallelisation is beneficial.

The mixing properties of the simulated MCMC chains are the same in \proglang{OpenBUGS} and \proglang{MultiBUGS}, because they use the same collection of underlying MCMC sampling algorithms.
Models with severe MCMC mixing problems in \proglang{OpenBUGS} are thus not resolved in \proglang{MultiBUGS}.
However, since \proglang{MultiBUGS} can speed-up MCMC simulation, it may be practicable to circumvent milder mixing issues by simply increasing the run length.

Several extensions and developments are planned for \proglang{MultiBUGS} in the future.
First, at present \proglang{MultiBUGS} requires the Microsoft Windows operating system.
However, most large computational clusters use the Linux operating system, so a version of \proglang{MultiBUGS} running on Linux is under preparation.
Second, \proglang{MultiBUGS} currently loads data and builds its internal graph representation of a model on a single core.
This process will need to be rethought for extremely large datasets and graphical models.

\section*{Acknowledgements}

\begin{sloppypar}
This work was supported by the UK Medical Research Council [programme codes MC\_UU\_00002/2 (RJBG), MC\_UU\_12023/21 (RT), MC\_UU\_00002/11 (DDA and AT)].
We are grateful to Chris Jewell, Sylvia Richardson and Christopher Jackson for helpful discussions of this work; to the Associate Editor and Reviewers for their insightful comments; and also to all contributors to the \proglang{BUGS} project upon which \proglang{MultiBUGS} is based.
\end{sloppypar}

\bibliography{bib}

\appendix

\section{Technical algorithmic details}
\label{sec:technical-details}

\subsection{Identifying conditionally independent parameters}
The following algorithm (called \code{find\_conditionally\_independent}) is used by \proglang{MultiBUGS} to identify sets of conditionally-independent parameters \(W_1, \dots, W_l \subseteq U\):

\begin{algorithmic}
\REQUIRE \(G = (E, V)\), a DAG; \(U\), a set of nodes (with identical topological depth)
\STATE \(l \leftarrow 1\)
\STATE \(M \leftarrow \varnothing\)
\WHILE {\(|U| > 0\)}
\FOR {\(u\) in \(U\)}
\IF {\(\ch_G(u) \cap M = \varnothing\)}
\STATE \(W_l \leftarrow W_l \cup \{u\}\)
\STATE \(U \leftarrow U \setminus \{u\}\)
\STATE \(M \leftarrow M \cup \ch_G(u)\)
\ENDIF
\ENDFOR
\STATE \(M \leftarrow \varnothing\)
\STATE \(l \leftarrow l + 1\)
\ENDWHILE
\ENSURE \(\{W_1, \dots, W_l\}\)
\end{algorithmic}

\subsection{Identifying parallelisable likelihoods}
Nodes for which the likelihood calculations should be partitioned across cores are identified using the following algorithm, called \code{find\_partial\_product\_parallel}:

\begin{algorithmic}
\REQUIRE \(G = (E, V)\), a DAG; \(C\), a number of cores; \(h\), a topological depth; \(\deepest\), the maximum topological depth in \(G\); \(T\), a computation schedule; \(r\), the current schedule row
\STATE \(U \leftarrow D^h_G\)
\STATE \(\overline{\ch} \leftarrow \text{mean}_{v \in \stochastic_{G}}|\ch_G(v)|\)
\FOR {\(u\) in \(U\)}
\IF {\(|\ch_G(u)| > 2 \times \overline{\ch}\) \OR \(\deepest = 1\)}
\STATE \(r \leftarrow r + 1\)
\FOR {\(c\) in \(1\) to \(C\)}
\STATE \(T_{rc} \leftarrow u\)
\ENDFOR
\STATE \(U \leftarrow U \setminus \{u\}\)
\ENDIF
\ENDFOR
\ENSURE \(\{T, U, r\}\)
\end{algorithmic}

\subsection{Building a computation schedule}
The overall algorithm for allocating compution to cores is as follows:

\begin{algorithmic}
\REQUIRE \(G = (E, V)\), a DAG; \(C\), a number of cores
\STATE Initialise \(T\), a table with \(C\) columns
\STATE \(r \leftarrow 0\)
\STATE \(\deepest \leftarrow \max_{v \in \stochastic_{G}}d_{G}(v)\)
\FOR {\(h\) in \(\deepest\) to \(1\)}
\STATE \(\{T, U, r\} \leftarrow \text{\code{find\_partial\_product\_parallel}}(G, C, h, \deepest, T, r)\)
\STATE \(\{W_1, \dots, W_l\} \leftarrow \text{\code{find\_conditionally\_independent}}(G, U)\)
\FOR {\(i\) in \(1\) to \(l\)}
\STATE \(c \leftarrow 0\)
\FOR {\(j\) in \(\max_{w \in W_i}|\ch_G(w)|\) to \(1\)}
\FOR {\(x\) in \(\{w \in W_i : |\ch_G(w)| = j\}\)}
\IF {\(c \bmod{C} = 0\)}
\STATE \(r \leftarrow r + 1\)
\STATE \(c \leftarrow 0\)
\ENDIF
\STATE \(T_{r(c+1)} \leftarrow x\)
\STATE \(c \leftarrow c + 1\)
\ENDFOR
\ENDFOR
\ENDFOR
\ENDFOR
\ENSURE \(T\)
\end{algorithmic}

\end{document}